\documentclass{article}

\usepackage{arxiv}

\usepackage[utf8]{inputenc} 
\usepackage[T1]{fontenc}    
\usepackage[hidelinks]{hyperref}       
\usepackage{url}            
\usepackage{booktabs}       
\usepackage{amsfonts}       
\usepackage{nicefrac}       
\usepackage{microtype}      
\usepackage{graphicx}
\usepackage{doi}

\usepackage{amsmath}
\usepackage{amssymb}
\usepackage[mathcal]{euscript}

\usepackage{bm}
\usepackage{epsfig}
\usepackage{wrapfig}
\usepackage{color}
\usepackage{xcolor}
\usepackage{cite}
\usepackage{amsthm}

\usepackage{enumitem}

\usepackage{algorithm}
\usepackage{algpseudocode}

\algnewcommand{\algorithmicand}{\textbf{ and }}
\algnewcommand{\algorithmicor}{\textbf{ or }}
\algnewcommand{\OR}{\algorithmicor}
\algnewcommand{\AND}{\algorithmicand}
\algnewcommand{\var}{\texttt}

\usepackage{hyperref}
\usepackage{cleveref}
\crefname{algorithm}{algorithm.}{algorithms.}

\usepackage{chngcntr}
\usepackage{apptools}

\title{A Novel Phase-Noise Module for 
the QUCS Circuit Simulator. Part I : the Periodic Steady-State.}

\author{Torsten Djurhuus \thanks{The authors are
with the Institute of Physics, Goethe University of Frankfurt am
Main, Max-von- Laue-Strasse 1, 60438, Frankfurt am Main.
(correspondence e-mail:
t.djurhuus@physik.uni-frankfurt.de).} \\
  Goethe-University Frankfurt\\
  \texttt{t.djurhuus@physik.uni-frankfurt.de} \\
  \And
  Viktor Krozer \\
  Goethe-University Frankfurt \& \\
  Ferdinand-Braun-Institut, Leibniz \\
  Institut für Höchstfrequenztechnik\\ 
  \texttt{krozer@physik.uni-frankfurt.de}\\
}

\renewcommand{\headeright}{\textit{arXiv} Preprint}
\renewcommand{\undertitle}{\textit{arXiv} Preprint}

\hypersetup{ pdftitle={An Oscilator Phase-Noise Calculator for 
QUCS Part I : the Steady-State.}, pdfsubject={eess.SY, math.CA,
math.DS}, pdfauthor={Torsten Djurhuus, Viktor Krozer},
pdfkeywords={steady-state simulation, numerical analysis, circuit simulation, 
circuit analysis, autonomous circuits, oscillators,
QUCS simulator, QUCS-S simulator.}, }

\begin{document}

\maketitle

\begin{abstract}
The paper discusses work done to expand and extend 
the capabilities of the open-source QUCS 
circuit simulator through the implementation 
of a computationally efficient time-domain steady-state analysis module, supporting simulation 
of autonomous circuits. To our knowledge, this represents the first time such 
an analysis module has been implemented in the QUCS environment. Hitherto, 
the only available option was a harmonic-balance module which was 
strictly limited to non-autonomous (driven) 
circuits. The research has several important 
scientific and industrial applications in the area 
of large-signal steady-state analysis of 
autonomous circuits \emph{e.g.} free-running and coupled oscillator 
circuit networks. The reported results will have great impact \emph{w.r.t.} 
analyzing, synthesizing and optimizing oscillatory behavior of 
various important industrial circuits and systems. The developed tool, 
furthermore, introduces support for simulating 
noise performance of circuits operating under 
large-signal conditions. This paper is the 
first part of a two-part series documenting the implementation 
of a novel (coupled)-oscillator 
phase-noise simulator engine in the QUCS environment.  
The goal of this undertaking is the advancement of the open-source QUCS project 
towards becoming a viable competitor to the 
commercial simulators currently on the market. 
\end{abstract}


\section{Introduction}
\label{sec0}

Efficient and rigorous modelling tools for analyzing, synthesizing and optimizing 
the large-signal Periodic Steady-State (PSS) of non-linear 
driven and autonomous circuits play a critical role 
in the design cycle of various circuits with 
important applications in \emph{e.g.} modern 
communication and remote-sensing systems.
Given the ever-increasing complexity of contemporary device models, 
and scale of modern circuit schematics, such work necessitates 
using an Electronic-Design-Automation (EDA) simulation program.  
\par
The work described herein considers the open-source Quite Universal Circuit Simulator (\texttt{QUCS}) 
\cite{QUCS_techreport2005,QUCS_report2007}, published under 
the GNU General Public License (GPL). 
The \texttt{QUCS} project was initiated in 2005 
\cite{QUCS_techreport2005} and has since gone through several development cycles, 
updates and tests resulting in a robust and computationally efficient 
simulator for analysis of large-scale digital and analog circuits 
including an accessible Graphical User Interface (GUI) for schematic capture. 
The distribution comes with several important device models built-in and includes 
an interface for both the \texttt{Verilog} and \texttt{VHDL} hardware 
description languages. The \texttt{QUCS} distribution contains two main 
modules : the simulation engine, known as \texttt{qucsator}, and the 
GUI application for schematic capture. They communicate through a text-file based 
netlist interface powered by a flex/yacc interpretation application. Herein, 
a choice was made to use the GUI 
application from the open-source (GPL) \texttt{QUCS{-}S} distribution \cite{Brinson2015,Brinson2016,Brinson2017} 
instead of the original \texttt{QUCS} GUI program.
The \texttt{QUCS{-}S} distribution represents a \texttt{QUCS} clone project, 
inheriting various modules from the original package while also 
adding a new GUI application \& external SPICE engine 
compatibility. It is furthermore able to directly interface 
with the original \texttt{qucsator} engine module. 
The package heritage structure described here was chosen after 
facing various practical problems installing 
the original \texttt{QUCS} GUI application on our workstation. 
\par
Unlike commercial counterparts, such as \emph{e.g.} 
Keysight-ADS\textsuperscript{\textcopyright} or
Cadence SpectreRF\textsuperscript{\textcopyright}, the \texttt{QUCS} package 
currently has no option for steady-state 
analysis of autonomous circuits (\emph{e.g.} free-running oscillators). 
A large-signal harmonic-balance (HB) simulator is 
included in the distribution, but this 
analysis option is limited to non-autonomous 
circuits \emph{e.g.} driven RF power amplifiers. 
Autonomous circuits, such as free-running oscillators/clocks, 
injection-locked \& phase-locked systems, bilaterally coupled oscillators 
\emph{etc.}, represent critically important circuit modules 
in modern communication and remote-sensing systems. The design of these applications 
necessitate robust numerical analysis tools for proper 
circuit development and design. Specifically, a PSS simulation tool is 
fundamentally important in-order to estimate, analyze, synthesize and optimize autonomous 
circuits \emph{w.r.t.} output signal power, steady-state frequency and 
stability of the large-signal solution. An option for simulating the PSS is 
furthermore required in-order to evaluate the large-signal noise-response 
\cite{djurhuus2005,djurhuus2005_2,djurhuus2006,djurhuus2009,djurhuus2021,djurhuus22} of (coupled) 
oscillator and clock circuit modules which are ubiquitous in modern electronic communication 
systems.
\par
The \texttt{QUCS}/\texttt{QUCS-S} project clone, described here,
is referred to as the \texttt{QUCS-COPEN} (working title) 
project/distribution/package \emph{etc.} with the acronym \texttt{COPEN} 
standing for Coupled Oscillator 
PhasE-Noise. The \texttt{QUCS-COPEN} project is currently a work-in-progress and, 
as the name suggest, involves the 
implementation of a phase-noise (PNOISE) analysis tool. This planned 
PNOISE engine module, which is based on a novel model developed by the authors in \cite{djurhuus22}, 
requires a numerically calculated PSS solution and hence depend directly upon the engine 
API developed herein. The implementation of the \texttt{QUCS-COPEN} PNOISE module 
is discussed in part 2 of this paper-series.

\subsection{A Note on Alternative Open-Source Engines : \texttt{qucsatorRF}, \texttt{GnuCap}, 
\texttt{ngspice} \& \texttt{Xyce}.}
\label{sec0:sub0}

The \texttt{QUCS-S} software package \cite{Brinson2015,Brinson2016,Brinson2017} 
ships with its own \texttt{qucsator} engine clone called \texttt{qucsatorRF}. 
From a quick initial inspection, this package looks very similar to the original. 
The \texttt{qucsatorRF} engine, 
being a clone of the \texttt{qucsator} engine, also 
does \underline{not} contain a PSS module for autonomous circuit analysis (see \texttt{qucsator} 
discussion above). A choice was made to exclusively work with the original 
\texttt{qucsator} engine for the \texttt{QUCS-COPEN} project.   
\par
Both the \texttt{QUCS} and \texttt{QUCS-S} packages include the 
option to link to other non-\texttt{QUCS}, \texttt{SPICE} based, open-source engines. The \texttt{QUCS} suite includes an interface to 
the \texttt{GnuCap} package\cite{GnuCapManual} whereas \texttt{QUCS-S} links to 
both the \texttt{ngspice} \cite{ngspiceManual} and \texttt{Xyce} \cite{XyceManual2025} engines. 
According to the latest published manual, the 
\texttt{GnuCap} API does not currently support any type of large-signal 
steady-state analysis \cite{GnuCapManual}. The \texttt{Xyce} simulator does support PSS 
analysis of driven/non-autonomous circuits (similarly to the \texttt{qucsator} \& 
\texttt{qucsatorRF} engines) but, importantly, \underline{does not} support PSS analysis of 
autonomous circuits \cite[section 7.7, second paragraph]{XyceManual2025}.
\par
The latest \texttt{ngspice} manual release does indeed talk about a PSS 
option for autonomous circuits \cite[section 11.3.12]{ngspiceManual}.
Importantly, this entry emphasizes (first line) : 
\emph{"Experimental code, not yet made publicly available."} Very 
little information exists \emph{w.r.t.} this 
PSS algorithm which, as far as the authors can tell, 
is only discussed in two sets of presentation slides \cite{Lanutti2011,Lanutti2012} 
and very briefly in the paper \cite{Grabinski2014}.
In the slides of \cite{Lanutti2011,Lanutti2012} the algorithm is referred to as 
the \emph{transient shooting} (TRAN-SHOOT) method. Despite its name, 
this algorithm seems to have very little in common with the classical shooting 
method (PSS-SHOOT) \cite{Aprille1972,Kundert1995,Kundert1997}.  
The PSS-SHOOT algorithm, utilized herein (see discussion below in \cref{sec1}),  
defines the PSS as the zero of non-linear system of equations. 
This represents a clearly stated problem with a well-understood 
solution procedure \emph{i.e.} numerical multidimensional root finding. 
On the other hand, the lack of documentation, makes it very difficult 
to gauge the performance and convergence properties of the experimental TRAN-SHOOT  
method. From our perspective, 
several open questions exist \emph{w.r.t.} the TRAN-SHOOT method, 
such as \emph{e.g.} for what set of initial 
conditions (basin of attraction) is convergence guaranteed?, 
what is the rate of convergence within this basin? \emph{etc.} Note, that 
these are all questions which can be answered (qualitatively at-least) 
in the context of the PSS-SHOOT algorithm employed herein. Use of the \texttt{ngspice} 
API is further complicated by its \texttt{SPICE 3} heritage. From \cite{Grabinski2014}, 
this entails, among other things, that it is not possible to sort 
circuit elements based on their component class. This type of operation is, however, 
required in-order to implement the PNOISE module discussed above. 
Due to the circumstance described here, a decision was made to implement our
own in-house PSS tool for the \texttt{QUCS-COPEN} project.

\subsection{Outline of paper}

Below, various theoretical and practical aspects \emph{w.r.t.} the 
\texttt{QUCS-COPEN} PSS simulator engine are discussed. 
\Cref{sec1,sec1:sub1,sec1:sub2} 
reviews the theory behind the implemented algorithm 
whereas \cref{sec1:sub3,sec1:sub4} discusses the various practical issues related to the actual 
implementation \& integration of the tool into the \texttt{QUCS-COPEN} package. 
The text here will, in places, contain brief discussion of C++ code \cite{Stroustrup2013} 
(the \texttt{QUCS} program language) 
including the use of some C++ notation and keywords. In \cref{sec2}, 
the new \texttt{QUCS-COPEN} 
PSS tool will be demonstrated on three 
example oscillator circuits. Examples are restricted to autonomous (clock) circuits as this 
aligns with the topic of this paper-series, \emph{i.e.} 
(coupled)-oscillator phase-noise analysis. 
Then in \cref{sec2:sub1} the results produced by the developed PSS tool 
are validated through a comparison study w/ the commercial Keysight-ADS\textsuperscript{\textcopyright} 
EDA HB simulator. 
The section also contains a detailed discussion of the convergence 
properties of the implemented tool. Finally, \cref{sec3} summarizes 
the obtained results and outlines 
future work \& projects, currently in the pipeline \emph{w.r.t.} 
to the \texttt{QUCS-COPEN} project.

\section{Background Theory \& PSS Module Implementation.}
\label{sec1}

The circuit equations are written using 
standard Modified-Nodal-Analysis (MNA) notation

\begin{equation}
\dot{q}(x(t)) + i(x(t)) + s(t) = 0
\label{sec1:eq1}
\end{equation}

where, $x(t) : \mathbb{R} \to \mathbb{R}^n$, represents the circuit state, 
assuming an $n$-dimensional system, with vectors, 
$q(x),i(x) : \mathbb{R}^n \to \mathbb{R}^n$, holding the reactive and resistive contributions, 
respectively, and $s(t) : \mathbb{R} \to \mathbb{R}^n$ representing the 
contributions of independent sources. Calculating the solution to 
\cref{sec1:eq1} represents an Initial-Value-Problem (IVP). 
The Linear-Response (LR) equations, generated from \cref{sec1:eq1}, have the form

\begin{equation}
d(C(t)\delta x(t))/ dt + G(t)\delta x(t) = 0 
\label{sec1:eq1b}
\end{equation}

where, $\delta x(t) : \mathbb{R} \to \mathbb{R}^n$, is the LR state vector, 
whereas, 
$C(t) {=} dq(s)/ds,G(t) {=} di(s)/ds\bigl|_{s=x(t)} : 
\mathbb{R} \to \mathbb{R}^{n\times n}$, are two time-dependent $n$-dimnesional Jacobian matrices governing the LR system dynamics.
\par
The PSS, $x_s(t) : \mathbb{R} \to \mathbb{R}^n$, 
represents a special solution of \cref{sec1:eq1}, subject to the 
Boundary-Condition (BC)

\begin{equation}
x_s(t_0) = x_s(t_0 + T_0) \Leftrightarrow  x_s(t_0) - x_s(t_0 + T_0) = 0
\label{sec1:eq2}
\end{equation}

with, $t_0 > 0$, being some offset and, $T_0>0$, the period of the PSS. 
\Cref{sec1:eq1}, subject to 
the \cref{sec1:eq2}, then no longer constitutes an IVP but instead is classified 
as a (2-point) Boundary-Value-Problem (BVP). How to calculate the 
solution to such a BVP is not immediately obvious. It might seem possible to simply integrate \cref{sec1:eq1} until 
all transients have died out and then represent the PSS in-terms of this asymptotic solution. This brute-force approach is, however, entirely unworkable 
practice\footnote{Electrical circuits generally 
represent stiff dynamical systems w/ widely separated time-constants which implies 
long transient integration time + small/narrow time-steps. Furthermore, the precision of a solution 
obtained using this approach would not suffice for subsequent sensitivity or 
noise-analysis.}.
\par
The \texttt{QUCS-COPEN} PSS calculator routine, documented herein, is designed
around a kernel procedure modelled on the PSS-SHOOT algorithm, 
first described in \cite{Aprille1972,Aprille1972_2}, and briefly 
reviewed below in \cref{sec1:sub2}. 
Over the years, considerable research has gone into improving 
this time-domain methodology (see \emph{e.g.} \cite{Kundert1995,Kundert1997,
Bittner2018,Parkhurst2002,Kakizaki1985} 
for reviews) and the scheme has been employed in various important research and industrial projects (see \emph{e.g.} 
Cadence SpectreRF\textsuperscript{\textcopyright} documentation). It 
represents a variant of the general \emph{shooting-method} algorithm.

\subsection{The Shooting-Method.}
\label{sec1:sub1}

Consider the dynamical system in \cref{sec1:eq1} on 
the interval $t\in [t_a,t_b]$ subject to some unspecified BC. 
The solution procedure involves meshing 
interval into $K$ sub-intervals $t_a = t_1< t_2 \dots < t_{K+1} = t_b$, 
and then defining \cite{Parkhurst2002}

\begin{equation}
G^q(x(t_i;s_{i-1}),s_j) = 0 \,\, , \quad  i=1,2\cdots ,K+1 \,, \,\, j,q=1,2,\cdots ,K
\label{sec1:eq3}
\end{equation}

where $G^q : \mathbb{R}^n \times \mathbb{R}^n  \to \mathbb{R}^n$ 
is a series of $K$ maps enforcing continuity at the mesh boundaries plus the 
BC at the endpoints, $x(t)$, is the state-vector (see \cref{sec1:eq1}) whereas, 
$s_k$, denotes the initial-value at mesh boundary $t_{k}$. For, $K>1$, the procedure 
is referred to as \emph{multiple-shooting} \cite{Parkhurst2002} 
whereas, $K=1$, describes the \emph{single-shooting} scenario\cite{Aprille1972,Aprille1972_2}. 
Below, only single shooting is considered w/ the mesh $t_a = t_1 < t_2 = t_b$.  At this point, a reduced notation 
is adopted in-order to keep expressions as simple as possible\footnote{
The initial value at $t_1 = t_a$ is represented by the symbol $x_0$ \emph{i.e.} 
$s_1 \equiv x_0$ and thus $x(t_2;s_1) \equiv x(t_2;x_0)$. Given the 
interval, $\tau = t_b-t_a$, the 
solution at $t_2=t_b$ is written $x(t_2;x_0) \equiv x_{\tau}(t_1,x_0)$. 
Finally, the absolute time-variable, $t_1$, is eliminated from the expression, 
which leaves $x_{\tau}(x_0)$. Note, this expression \underline{implicitly} contains 
the absolute time-variable index, $t_1$, through the inclusion of $x_0$ which, by definition,
is tied to this time-point. Hence, although not explicitly stated, the expression, 
$x_{\tau}(x_0)$, includes absolute time-dependence and thus 
represents a valid notation for both 
autonomous and non-autonomous systems.\label{sec1:foot2}}. The solution 
at, $t_2 = t_b$, is then written, $x_{\tau}(x_0)$, with $\tau = t_b-t_a$ 
whereas, $x_0$, represents the 
initial value at $t_1=t_a$ (see \cref{sec1:foot2} for further explanation). 
Applying this new notation, the single-shooting variant of \cref{sec1:eq3} is 
written

\begin{equation}
F(x_0) =  G^1(x_{\tau}(x_0), x_0 ) = 0
\label{sec1:eq4}
\end{equation}

The expression in \cref{sec1:eq4} transforms the original BVP into
an IVP combined w/ the added problem of finding a root of an $n$-dimensional multivariate map. 
This is a well understood problem w/ several established solution procedures. 
Using \emph{e.g.} a Newton-Raphson (NR) solver approach, 
w/ initial guess, $x_{0}^{(0)}$, (in the basin of attraction) 
a solution can be iterated as follows  

\begin{equation}
x_0^{(l+1)} = x_0^{(l)}- \bigl[J^{(l)}\bigr]^{-1}F\bigl(x_0^{(l)}\bigr)
\label{sec1:eq5}
\end{equation}

where, $x_0^{(l)}$, is the $l$th iterate of the solution, 
$F : \mathbb{R}^n \to \mathbb{R}^n$ is given in \cref{sec1:eq4}, 
and $J^{(l)} \in \mathbb{R}^{n\times n}$ is the Jacobian 
of this map at $x_0^{(l)}$

\begin{equation}
J^{(l)} = \partial F(x_0^{(l)}) /\partial x_0^{(l)} = 
\Gamma_{\text{\tiny A}}^{(l)} \Phi_{\tau}^{(l)} + \Gamma_{\text{\tiny B}}^{(l)} 
\label{sec1:eq5b}
\end{equation}

Here $\Gamma_{\text{\tiny A}}^{(l)},\Gamma_{\text{\tiny B}}^{(l)} \in \mathbb{R}^{n\times n}$ are the $l$th iterate 
Jacobians of $G^1 : \mathbb{R}^n \times \mathbb{R}^n \to \mathbb{R}^n$ \emph{w.r.t.} 
its arguments and, $\Phi_{\tau}$, is the sensitivity matrix

\begin{equation}
\Phi_{\tau}^{(l)} = \partial x_{\tau}(x_0^{(l)})/ \partial x_0^{(l)}
\label{sec1:eq6}
\end{equation}

which can be calculated by integrating the LR equations in \cref{sec1:eq1b} 
in parallel with the full solution, calculated through \cref{sec1:eq1}.

\subsection{The PSS-SHOOT Method.}
\label{sec1:sub2}

The PSS-SHOOT method represents a special version the general 
algorithm discussed above. From \cref{sec1:sub1,sec1:eq2}, we
consider the mesh $t_a=t_0$, $t_b = t_0 + T_0$ and 
$\tau = t_1-t_0 = T_0$. From \cref{sec1:eq4}, the PSS problem is then formulated

\begin{equation}
F_P(x_0) = G^1(x_T(x_0),x_0) = x_0 - x_T(x_0) = 0
\label{sec1:eq7b}
\end{equation}

From \cref{sec1:eq5}, in \cref{sec1:sub1}, the root in \cref{sec1:eq7b} is found 
through iteration

\begin{equation}
x_{0}^{(l+1)} = x_{0}^{(l)}- \bigl[J_P^{(l)}\bigr]^{-1}\bigl(x_{0}^{(l)} - x_{T}^{(l)}\bigr) ,
\label{sec1:eq8}
\end{equation}

where the Jacobian, $J_P^{(l)}$, follows from  \cref{sec1:eq5b} subject to the 
PSS map in \cref{sec1:eq7b}. From inspection of \cref{sec1:eq7b}, 
$\Gamma_{\text{\tiny A}} = -I_n$ and $\Gamma_{\text{\tiny B}} = I_n$, 
(see discussion in \cref{sec1:sub1}) where $I_n \in \mathbb{R}^{n\times n}$ is the $n$-dimnesional identity matrix. Inserting 
into \cref{sec1:eq5b} gives

\begin{equation}
J_P^{(l)} = I - \Phi_T^{(l)}
\label{sec1:eq9}
\end{equation}

with, $\Phi_T^{(l)} = \partial x_{T}(x_0^{(l)})/ \partial x_0^{(l)}$, 
being the $l$th iterated sensitivity matrix (see \cref{sec1:eq6} above).

\subsubsection{The Autonomous Case.}
\label{sec1:sub2:sub1}

The text above assumes that the frequency of operation, $f_0=1/T_0$, is a known fixed parameter. 
For autonomous circuits, \emph{e.g.} free-running oscillators, this is of-course 
not the case. Below, the augmented PSS representation \& operators, 
for the autonomous case, will be derived. In-order to limit notational 
complexity, the iteration index, $l \in \mathbb{Z}_{\geq 0}$, (see \emph{e.g.} \cref{sec1:eq8}) 
will be neglected. From the discussion in \cref{sec1:sub1,sec1:sub2} above, 
it should be clear how to re-introduce this index into the derived expressions.
\par
The period, $T_0$, is now a system variable and thus must be included in the 
state-vector. The augmented PSS 
state-vector, $\tilde{x}_0 : \mathbb{R}^{n+1}$ is then written 

\begin{equation}
\tilde{x}_0 = \begin{bmatrix} x_0 & T_0 \end{bmatrix}^{\top}
\label{sec1:eq12}  
\end{equation}

With $n{+}1$ variables, the $n$-dimensional map in \cref{sec1:eq7b} is now under-determined. 
To fix this issue, an additional equation, $\alpha^{\top}x_0 = 0$, 
is added to the system. Here, $\alpha \in \mathbb{R}^n$ is a constant 
$n$-dimensional vector and the introduced equation hence simply fixes of the phase of the PSS solution. 
The augmented PSS map is then written  

\begin{equation}
\tilde{F}_P(\tilde{x}_0) =  \begin{bmatrix} F_P(\tilde{x}_0) \,\,\,\, \alpha^{\top}x_0  \end{bmatrix}^{\top}    
\label{sec1:eq14}
\end{equation}

where, $F_P$, is the map introduced above in \cref{sec1:eq7b}. 
The augmented Jacobian, defined as the derivative of the map in \cref{sec1:eq14}, 
is then written from inspection

\begin{equation}
\tilde{J}_P = \begin{bmatrix} J_p & \Psi_T  \\
\alpha^{\top} & 0
\end{bmatrix}  =    \begin{bmatrix} I - \Phi_T & \Psi_T \\
\alpha^{\top} & 0
\end{bmatrix}
\label{sec1:eq15}  
\end{equation}

where, $J_P$, is the original PSS Jacobian, $\Phi_T$, is the sensitivity matrix (see \cref{sec1:eq9}) and 
the vector, $\Psi_T \in \mathbb{R}^n$, holds the components of vector, 
$-\partial x_{T}(x_0) / \partial T_0 \in \mathbb{R}^n$. 
Employing the augmented operators, \cref{sec1:eq12,sec1:eq14,sec1:eq15}, 
(w/ re-introduced iteration index $(l)$) in \cref{sec1:eq8} of \cref{sec1:sub2}, 
the PSS-SHOOT algorithm is readily extended to include autonomous circuits.

\subsection{Implementing the \texttt{QUCS-COPEN} PSS Module.}
\label{sec1:sub3}

\begin{algorithm}
\caption{\texttt{QUCS-COPEN} PSS Module Kernel.}\label{sec1:alg1}
\textbf{Parameters :} $\texttt{Tper}, \texttt{Tstab}, \texttt{MaxItr}, \texttt{EpsMax}$ 
\Comment{\texttt{Tper} : init. period guess, \texttt{Tstab} : stabilization time.} \\
\textbf{Input :} $x_{\text{DC}}$  \Comment{calculated DC solution.}\\
\textbf{Output Datasets :} init. tran. (Xt), PSS (Xp), abs. spec. (Xpa)
\Comment{DS file ext. w/ X=[V/I], Volt./Current.}
\begin{algorithmic}[1]
\Require $\texttt{Tper}>0,\texttt{Tstab}\geq 10\times \texttt{Tper}, \text{MaxItr} \geq 10, \texttt{EpsMax} \leq 10^{-6}$
\Ensure $\texttt{isNum}(x_{\text{DC}}) = \texttt{true}$ \Comment{check : $x_{\text{DC}}$  is valid DC solution.}
\State $x_{\text{\tiny init}} \gets \texttt{TranInit}(0,\texttt{Tstab};x_{\text{DC}})$ 
\label{sec1:alg1:tranCall1}
\State $t_0 \leftarrow \texttt{Tstab}$, $T_0 \leftarrow \texttt{Tper}$
\State $x_0^{(0)} \gets x_{\text{\tiny init}}$
\State $x_T^{(0)} \gets \texttt{TranPss}(t_0,t_0+T_0;x_0^{(0)})$
\Comment eval PSS, $x_s^{(l=0)}(t)$ : $x_s^{(l)}(t_0) {=} x_0^{(l)}$, $x_s^{(l)}(t_0+T_0){=}x_T^{(l)}$ 
\label{sec1:alg1:tranCall2}
\State $l \leftarrow 0$, $\texttt{done} \gets \texttt{false}$
\While{ $\texttt{done} = \texttt{false} \AND l \leq \text{MaxItr}$ }
\State $G^{(l)}(t), C^{(l)}(t) \gets \texttt{collectGC}(x_s^{(l)}(t))$ \label{sec1:alg1:collectGC}
\Comment{eval $G^{(l)}(t),C^{(l)}(t)$ along path $x_s^{(l)}(t)$ (see \cref{sec1:eq1b}). }
\State $\Phi_T^{(l)} \gets \texttt{Tran}_{\Phi}(t_0,t_0+T_0;x_0^{(l)})$
\Comment{calc. sensitivity matrix (see \cref{sec1:eq6,sec1:eq9}). \label{sec1:alg1:smatrix}}
\State $J_{\text{P}}^{(l)} \gets I - \Phi_T^{(l)}$
\State $F_P^{(l)} \gets x_0^{(l)}- x_T^{(l)}$
\State $x_{0}^{(l+1)} \gets x_{0}^{(l)} {-} \bigl[J_P^{(l)}\bigr]^{-1}F_P^{(l)}$
\State $x_T^{(l+1)} \gets \texttt{TranPss}(t_0,t_0+T_0;x_0^{(l+1)})$ 
\Comment generates $l{+}1$th iterated PSS, $x_s^{(l+1)}(t)$, (see \cref{sec1:alg1:tranCall2} comment) 
\label{sec1:alg1:tranCall3}
\State $\epsilon^{(l)} \gets \Vert x_T^{(l+1)} {-} x_0^{(l+1)} \Vert$
\If{$\epsilon^{(l)} <  \texttt{EpsMax}$}
    \State $\texttt{done} \gets \texttt{true}$ \Comment{PSS solution found, exit routine,}
\Else
    \State $l \gets l + 1$
\EndIf
\EndWhile
\State $x_s(t) \gets x_s^{(l+1)}(t)$ \Comment store iterated PSS solution. 
\end{algorithmic}
\end{algorithm}

The \texttt{QUCS-COPEN} PSS module kernel is briefly outlined in \cref{sec1:alg1}. 
As discussed above, this kernel is designed around the PSS-SHOOT algorithm 
reviewed above in \cref{sec1:sub2,sec1:sub2:sub1}. 
The module code resides within the new \texttt{psssolver} C++ class, 
which, in-turn, inherits directly from the existing \texttt{trsolver} class 
containing the \texttt{qucsator} transient engine implementation. 
For reasons discussed below, it was necessary to refactor 
and slightly rewrite, the original transient engine code in \texttt{trsolver}. 
Care was taken to maintain integrity of the original \texttt{qucsator} API, inline with our goal of 
extending the simulator w/o changing or destroying 
the interface and/or already existing functionality.  
\par
In \cref{sec1:alg1} (\cref{sec1:alg1:tranCall1,sec1:alg1:tranCall2,sec1:alg1:tranCall3}) 
calls are made to transient engine subroutines $\texttt{TranInit}$ and $\texttt{TranPss}$, 
both which are member methods belonging to the \texttt{trsolver} base-class. 
Here, calls of the type $\texttt{Tran[Init/Pss]}(t_a,t_b,x_0)$ both generate 
paths, $x(t_a) \to x(t_b)$, by integrating the IVP in \cref{sec1:eq1}, subject 
to initial condition $x(t_a) = x_0$. Despite this similarity, 
the implementation of these two methods is notably different in-terms of \emph{e.g.} 
initialization of solution state arrays, class state variable histories, the unwieldy 
large set of \texttt{trsolver} protected class members 
(acting as quasi-global state variables), various solver mode variables 
\& switches, as well the invocation of allocation and clean-up/exit routines \emph{etc}. 
The original \texttt{qucsator} API was simply not written to handle 
this level of flexibility described here \emph{w.r.t.} different flavors of 
transient engine solver calls. This problem, in addition to a myriad of other 
implementation issues \emph{w.r.t} the \texttt{trsolver}->\texttt{psssolver} 
C++ inheritance architecture, triggered the refactoring of the QUCS-COPEN 
transient engine mentioned above. 
\par
The method, $\texttt{collectGC}$, called on \cref{sec1:alg1:collectGC} of \cref{sec1:alg1} 
is a \texttt{psssolver} subroutine responsible for collecting the Jacobian matrices, $C(t),G(t)$, 
introduced in \cref{sec1:eq1b}, along the 
path, $x_s(t)$, $t\in[t_0;t_0+T_0]$ 
(see discussion \emph{w.r.t.} \texttt{TranPss} above). 
This job is complicated somewhat by the fact that the 
\texttt{qucsator} implementation of the circuit equations does \underline{not} 
employ the standard MNA format shown in \cref{sec1:eq1,sec1:eq1b}. 
The \texttt{qucsator} circuit equations (transient analysis) are instead written \cite{QUCS_techreport2005}
$A(t) x(t) = I_{eq}(t) + s(t)$, where, 
$A(t) : \mathbb{R} \to \mathbb{R}^{n\times n}$ represents the 
Jacobian matrix and $x(t), s(t)$ again denote the 
system state/source-vectors (see \cref{sec1:eq1}). 
The vector function, $I_{eq}(t) : \mathbb{R} \to \mathbb{R}^n$, holds 
the \emph{equivalent source vector} which contains past Linear-Multistep (LMS) terms of reactive components, 
nonlinear resistive component contributions and artificial offset terms which are included to 
fix double counting issues \cite{QUCS_techreport2005}.
The two MNA representation discussed here (\cref{sec1:eq1} + the above) are 
obviously equivalent as they must produce the same set of KCL/KVL equations. 
From standard LMS theory, $A(t) = a_0C(t) + G(t)$, 
where the scalar, $a_0\in \mathbb{R}$, depends on the specific LMS method and order employed. 
The matrix, $G(t)$, represents the DC Jacobian matrix function which can 
be calculated using the existing \texttt{qucsator} API. Subtracting this matrix from the 
above expression for the Jacobian allows for the evaluation of 
$a_0C(t)$, and hence $C(t)$. The call to \texttt{psssolver} method, $\texttt{Tran}_{\Phi}$, 
on \cref{sec1:alg1:smatrix} of \cref{sec1:alg1} then proceeds to calculate 
the circuit sensitivity matrix by integrating the LR equation in \cref{sec1:eq1b} 
along the path, $x_s(t)$, using the matrix functions evaluated in $\texttt{collectGC}$. 
This LR method must follow the exact same LMS method \& order, 
time-step mesh, program state histories \emph{etc.} as the \texttt{trsolver} 
integration routine, $\texttt{TranPss}$, which generated $x_s(t)$.

\subsection{The GUI Environment \& Output Datasets.}
\label{sec1:sub4}

\begin{figure}[!h]
\begin{center}
\includegraphics[scale=0.75]{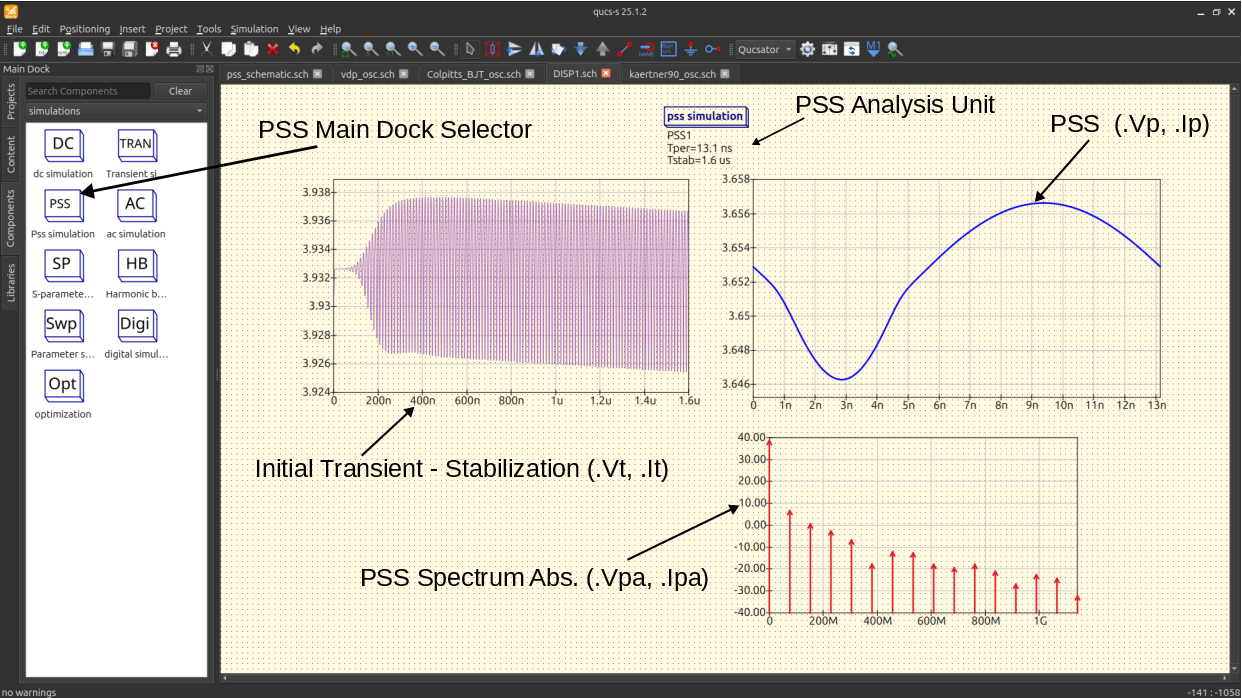}
\end{center}
\caption{The figure shows the \texttt{QUCS-COPEN} 
PSS simulation unit integrated into the 
\texttt{QUCS-S} GUI environment.  The analysis module is chosen 
from the main-dock and then placed on the schematic. 
The unit has two main parameters : \texttt{Tper} and \texttt{Tstab}, and the 
PSS precision is fixed in all experiments at \texttt{EpsMax}=$10^{-12}$ (see \cref{sec1:alg1})
The 3 plotted graphs represent the 3 types of output datasets generated by the 
PSS simulator (unspecified oscillator circuit \& output node). These are (w/ file extensions), 
\textcolor{purple}{purple graph} : initial/stabilization transient, 
(\texttt{.Vt}, \texttt{.It}), \textcolor{blue}{blue graph} : 
PSS time-domain solution, (\texttt{.Vp}, \texttt{.Ip}) and \textcolor{red}{red graph} : PSS solution, 
frequency-domain spectrum, absolute value, (\texttt{.Vpa}, \texttt{.Ipa}). which is plotted in \texttt{dBm} }
\label{sec1:fig1}
\end{figure}

As mentioned in \cref{sec0}, the choice was made to link \texttt{QUCS-COPEN} extended simulator engine  
to \texttt{QUCS-S} GUI environment. \Cref{sec1:fig1} illustrates
newly implemented PSS tool operating inside this application. The figure shows the 
result of running a PSS simulation for some 
unspecified oscillator circuit. The PSS simulator (currently) outputs 
3 datasets (see also \cref{sec1:alg1}) : the stabilization transient (file-extensions \texttt{.Vt,.It}), 
the PSS time-domain solution (file-extensions \texttt{.Vp,.Ip}) 
and the PSS absolute-value spectrum (file-extensions \texttt{.Vpa,.Ipa}). 
From the figure it is also seen that the 
PSS analysis module has two main parameters : 
\texttt{Tper} \& \texttt{Tstab}, which are described in the comments 
of \cref{sec1:alg1} (parameter section).

\begin{figure}[!h]
\begin{center}
\includegraphics[scale=0.725]{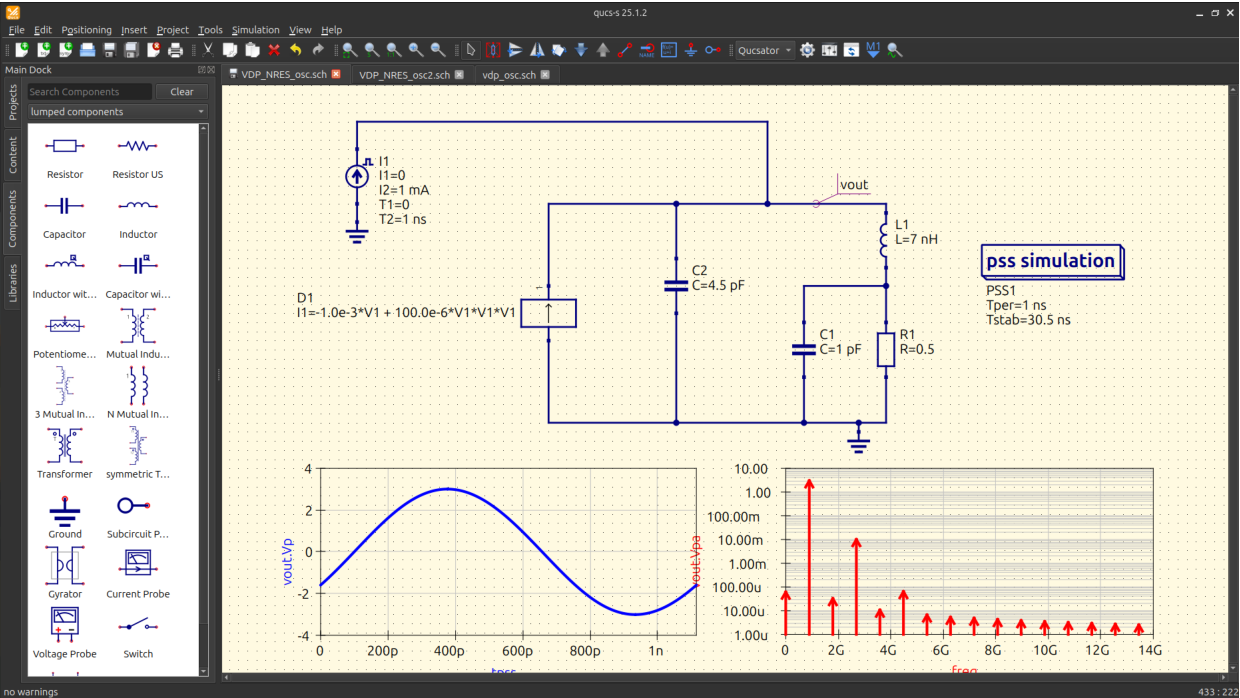}
\end{center}
\caption{The figure shows the \texttt{QUCS-COPEN} PSS calculator 
applied to a simple Van-der-Pol(VDP)-type oscillator first proposed in 
\cite[Fig. 3]{djurhuus22}. This circuit is referred to below as \texttt{OSC.\#1}.
All parameters are as in \cite{djurhuus22} 
(primary osc. parameter set). Two of the three available PSS datasets 
are plotted for the voltage node, \texttt{vout}, 
(stabilization transient data is not shown). Refer to \cref{sec1:fig1} 
caption for details.}
\label{sec2:fig1}
\end{figure}

\section{Example Oscillator Circuits.}
\label{sec2}

Below, the \texttt{QUCS-COPEN} PSS solver is applied to three different oscillator circuits. 
As far as the authors can tell, the tool demonstrated 
here represents the first successful implementation of such a module, applicable to simulation of nonlinear \underline{autonomous circuits}, 
into the \texttt{QUCS} ecosystem. 
\par
The first example considered is a simple Van-der-Pol equivalent 
circuit, shown \cref{sec2:fig1}. The oscillator negative 
conductance is implemented using a \texttt{QUCS} Equation-Defined-Device (EDD) \cite{QUCS_report2007} 
w/ the current/voltage characteristic 
$I_n = a_0 + a_1V + a_2V^2 + a_3 V^3$, where $a_0=a_2 = 0$ and 
$a_1 = -1.0\mathrm{mS}$, $a_3 = 100.0 \mathrm{\mu A/V^3}$. 
Main PSS parameters are set as : $\texttt{Tper} = 1.0 \mathrm{ns}$ and 
$\texttt{Tstab} = 30.5 \mathrm{ns}$. The PSS module calculates 
a solution with oscillation frequency $f_0 = 1/T_0 = 896.6 \mathrm{MHz}$. The figure 
plots the PSS solution and abs. PSS spectrum datasets, 
see also \cref{sec1:sub4,sec1:fig1} for details on datasets.

\begin{figure}[!h]
\begin{center}
\includegraphics[scale=0.725]{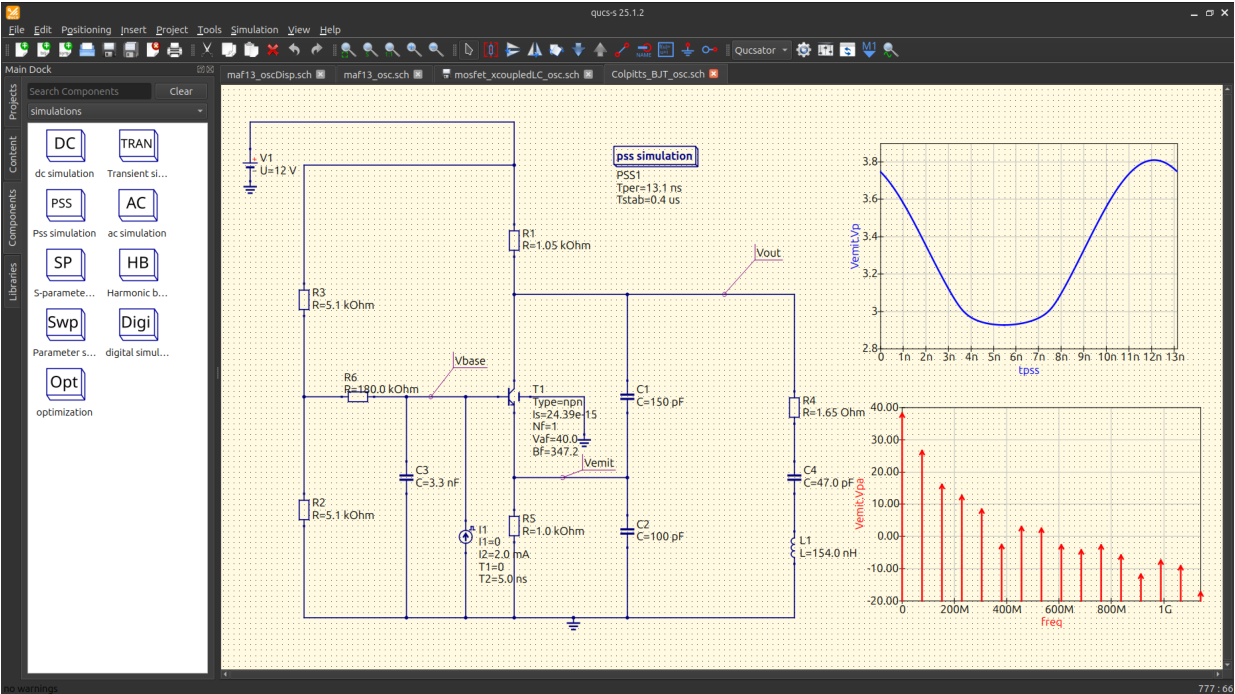}
\end{center}
\caption{ The simulator is applied to a 
BJT Colpitts oscillator taken from \cite{Kaertner1990}. All circuit parameters 
as in \cite{Kaertner1990}, w/ the following exceptions : 
resonator resistor $r_0  = 1.65 \mathrm{Ohm}$ (was $0.65 \mathrm{Ohm}$ in the paper) 
and the BJT nonlinear base resistor is set to zero due to an internal \texttt{qucsator} BJT model 
issue (see \cref{sec2:foot1}). This circuit is referred to below as \texttt{OSC.\#2}. 
The figures plot the PSS output datasets for the 
external BJT emitter node, referred to as, \texttt{Vemit}, in the schematic. See the captions of 
\cref{sec1:fig1,sec2:fig1} for details.}
\label{sec2:fig2}
\end{figure}

\begin{figure}[!h]
\begin{center}
\includegraphics[scale=0.725]{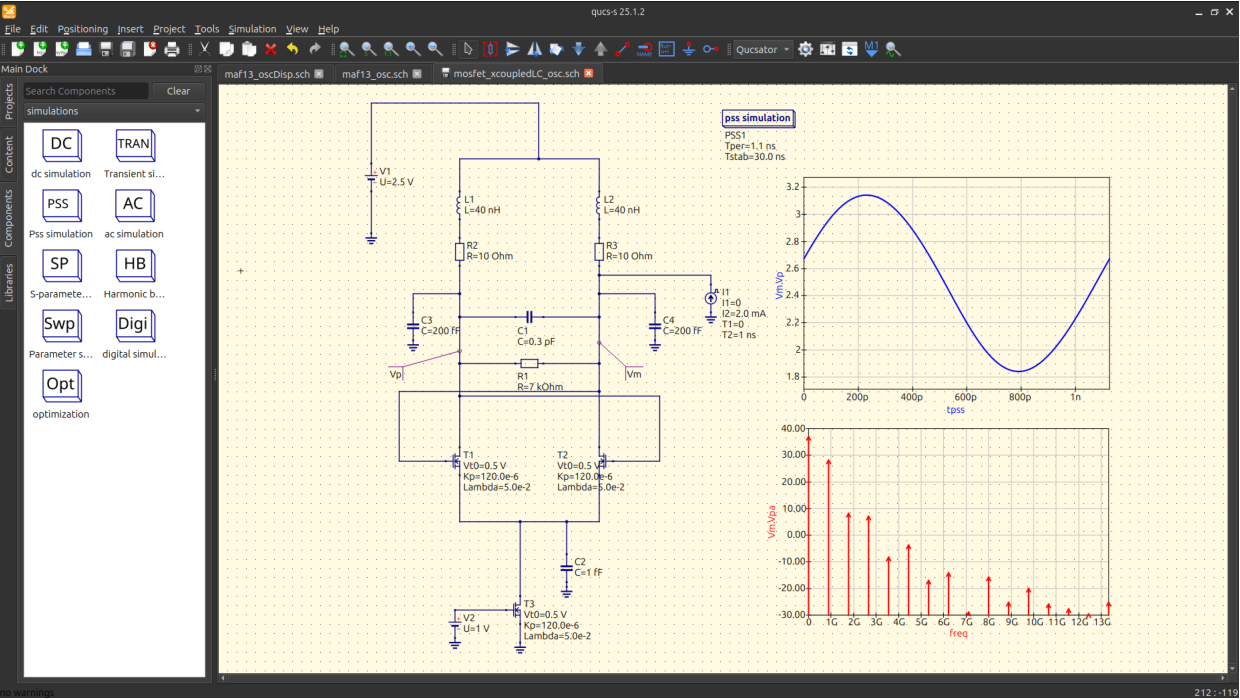}
\end{center}
\caption{Figure shows the \texttt{QUCS-COPEN} PSS simulation tool applied to a 
MOSFET cross-coupled LC-tank oscillator. The circuit is similar to the 
oscillator proposed in \cite[fig. 3]{Maffezzoni2013} with minor variations (see text). 
This circuit is referred to below as \texttt{OSC.\#3}.
The figures plot the PSS output datasets for the right-hand MOSFET device drain-node, 
referred to as, \texttt{Vm}, in the schematic.
See also \cref{sec1:fig1,sec2:fig1} for details. }
\label{sec2:fig3}
\end{figure}

\Cref{sec2:fig2} shows the same simulation setup 
as in \cref{sec2:fig1} but this time for a BJT Collpitts oscillator 
first proposed in the paper \cite{Kaertner1990}. 
Compared with the original oscillator design, the circuit in \cref{sec2:fig2} includes 
a few modifications. Firstly, 
the resistor in the series LCR resonator was changed from $0.65 \mathrm{Ohm}$ 
in \cite{Kaertner1990} to $1.65 \mathrm{Ohm}$ in the schematic shown here. 
Secondly, due to some internal \texttt{qucsator} BJT modelling issues (see 
\cref{sec2:foot1}), 
which will be fixed in future releases of \texttt{QUCS-COPEN}, 
it was found necessary to neglect the BJT nonlinear base-resistance 
contribution\footnote{After a rather lengthy debug session it was 
found that the \texttt{qucsator} code 
did not fully implement the Jacobian 
matrix contributions for the BJT base resistance. It only included 
the first-order contributions, \emph{i.e.} the 
resistor itself, but not derivatives of the nonlinear characteristic. 
This type of modelling seems to suffice for DC, transient simulation \emph{etc.}, 
however, the PSS calculation involves 
integrating the Jacobian (see \cref{sec1:sub3,sec1:alg1}),
alongside the actual solution, and this modelling issue 
was found to lead to numerical instability problems.
\label{sec2:foot1}
}.  Here, this internal 
component was cancelled by setting BJT model parameters 
$\texttt{Rbm} = \texttt{Rb} = 0.0 \mathrm{Ohm}$. All remaining 
contributions of the nonlinear BJT model are included in the simulation. 
The \texttt{QUCS-COPEN} PSS calculator finds a solution oscillating at, 
$f_0 = 76.07 \mathrm{MHz}$, and the output datasets 
are plotted for the external emitter node of the BJT device. Finally, \cref{sec2:fig3} shows 
the PSS module applied to a MOSFET cross-coupled pair, LC tank oscillator. 
The circuit is similar to the MOSFET oscillator proposed in 
\cite[fig. 3]{Maffezzoni2013} including identical device models. The circuit 
in \cref{sec2:fig3} differs from the original by the addition of inductor series 
resistances $r_s = 10.0\mathrm{Ohm}$. The simulator finds a solution oscillating 
at, $f_0 = 888.6 \mathrm{MHz}$, and the PSS output datasets are plotted.

\subsection{ Verification of the Obtained PSS Solutions \& Convergence Tests.}
\label{sec2:sub1}

\begin{table}
\centering
\includegraphics[scale=1.1]{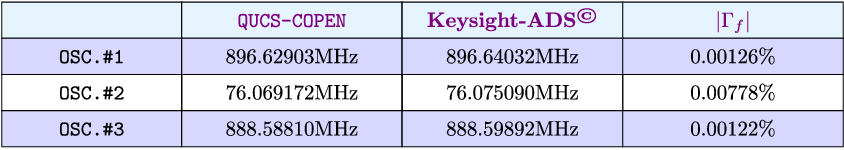}
\caption{The table lists PSS oscillator frequencies calculated using the \texttt{QUCS-COPEN} package and the commercial Keysight-ADS\textsuperscript{\textcopyright} 
simulator for circuits : \texttt{OSC.\#1} = VDP circuit in \cref{sec2:fig1}, 
\texttt{OSC.\#2} = BJT circuit in \cref{sec2:fig2} and \texttt{OSC.\#3} = MOSFET circuit in \cref{sec2:fig3}.  
The relative deviation between the calculated values are 
recorded by measure $\Gamma_f$ (see text in \cref{sec2:sub1}).}
\label{sec2:tab1}
\end{table}

\begin{figure}[!th]
\begin{center}
\includegraphics[scale=1.0]{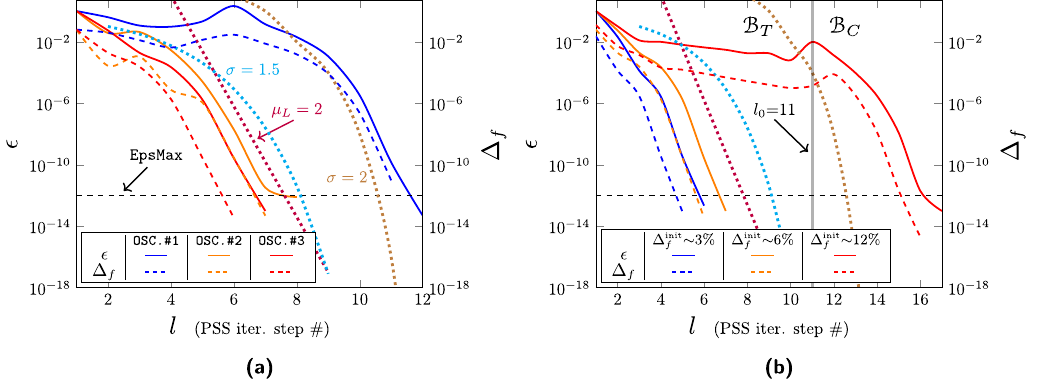}
\end{center}
\caption{Convergence measures, $\epsilon$ (solid line left y-axis), and  
$\Delta_f$ (dashed line, right y-axis), 
defined in \cref{sec2:eq1,sec2:eq2}, plotted as a function of the PSS iteration index 
$l$ (see \cref{sec1:sub1,sec1:sub2,sec1:alg1}) for different circuits and parameter configurations.  
Dotted curves represent sequences of varying convergence order $\sigma$ 
(see \cref{app1:sec1}) : 
$\sigma=2$ (brown \textcolor{brown}{\scalebox{1.5}{$\cdots$}}), $\sigma=1.5$ (cyan \textcolor{cyan}{\scalebox{1.5}{$\cdots$}}),
$\sigma=1$ (purple \textcolor{purple}{\scalebox{1.5}{$\cdots$}}) w/ linear convergence rate $\mu_L = 2$.  
(a) : PSS measures, $\epsilon,\Delta_f$, are plotted for each of the oscillators 
\texttt{OSC.\#1-3} (see \cref{sec2:tab1} caption  \& \crefrange{sec2:fig1}{sec2:fig3}) w/ initial conditions, 
$\Delta_f^{\text{\tiny init}} \sim 6\%$ 
and $K_{\text{\tiny STAB}} \sim 30{-}35$ (see \cref{sec2:eq3,sec2:eq4}). (b) : figure plots the PSS measures for circuit 
\texttt{OSC.\#3}, for 3 different initial conditions as specified by the 
parameter $\Delta_f^{\text{\tiny init}}$ and w/ $K_{\text{\tiny STAB}} \sim 30{-}35$.}
\label{sec2:fig4}
\end{figure}

Define the measure, $\Gamma_f = (f_0^{\text{\tiny QC}} -f_0^{\text{\tiny K-ADS}} )/f_0^{\text{\tiny QC}}$, 
where $f_0^{\text{\tiny QC}}$ and $f_0^{\text{\tiny K-ADS}}$ are the steady-state 
oscillator frequencies calculated using the \texttt{QUCS-COPEN} 
PSS tool and the 
commercial Keysight-ADS\textsuperscript{\textcopyright} HB simulator, respectively. 
Thus, $\Gamma_f$ records, for a given oscillator circuit, the deviation of PSS frequencies 
calculated applying these two very different EDA programs. \Cref{sec2:tab1} 
displays the calculated PSS frequencies and the resulting measure, $|\Gamma_f|$, 
for each of the three oscillators discussed above in \cref{sec2}. 
The maximum measured relative error is $|\Gamma_f| \sim 0.008\%$, or 
$|\Gamma_f| \sim 80 \mathrm{ppm}$ (parts-per-million). The results listed 
in \cref{sec2:tab1} serve to validate the developed \texttt{QUCS-COPEN} PSS tool.
\par
Two PSS convergence measures are defined

\begin{align}
\epsilon(l) &=  \Vert x_s^{(l)} - \bar{x}_s\Vert  \label{sec2:eq1}\\
\Delta f(l) &=  \Vert f_0^{(l)} - \bar{f}_0\Vert / \bar{f}_0 \label{sec2:eq2}
\end{align}

where $\epsilon(l) : \mathbb{Z}_{\geq 0} \to \mathbb{R}$, 
in \cref{sec2:eq1}, represents the precision of the $l$th 
iterated PSS solution, $x_s^{(l)}$, (see \cref{sec1:sub2,sec1:sub3} 
and \cref{sec1:alg1}) with $\bar{x}_s$ denoting the true PSS solution; 
\emph{i.e.} the numerically calculated PSS solution. Likewise, 
$\Delta f(l) : \mathbb{Z}_{\geq 0} \to \mathbb{R}$, 
in \cref{sec2:eq2}, measures the relative frequency precision of the $l$th iterated 
solution with $f_0^{(l)}$ and $\bar{f}_0$ denoting the frequency of the 
$l$th iterate and the true PSS frequency (\emph{i.e.} frequency of $\bar{x}_s$), 
respectively. To quantify initial conditions of experiments below, two further measures are 
defined

\begin{align}
\Delta_f^{\text{\tiny init}} &:= \Delta f(0) \label{sec2:eq3}\\
K_{\text{\tiny STAB}}  &:= \texttt{Tstab}\mod \bar{T}_{0} \label{sec2:eq4}
\end{align}

where $\Delta_f^{\text{\tiny init}}$ is the relative error 
of the initial PSS frequency guess and $K_{\text{\tiny STAB}}$ 
represents the stabilization interval (see \cref{sec1:alg1}) expressed 
to the nearest multiple of the PSS period $\bar{T}_{0} = 1/\bar{f}_0$.  
\par
\Cref{sec2:fig4} shows the results of a series of convergence 
experiments expressed in-terms of the measures defined above in \cref{sec2:eq1,sec2:eq2}. 
Specifically, \cref{sec2:fig4}.(a) plots these two measures 
as a function of the iteration index, $l \in \mathbb{Z}_{\geq 0}$, 
for each of the three oscillator circuits, \texttt{OSC.\#1-3}, shown in 
\crefrange{sec2:fig1}{sec2:fig3}, for a fixed set of initial conditions 
$(\Delta_f^{\text{\tiny init}},K_{\text{\tiny STAB}})$ 
whereas \cref{sec2:fig4}.(b) plots the measures for the \texttt{OSC.\#3} circuit 
(see \cref{sec2:fig3}), for a range of initial conditions, as 
described by $\Delta_f^{\text{\tiny init}}$ in \cref{sec2:eq3}.

\subsubsection{The Convergence Zone $\mathcal{B}_C$.}
\label{sec2:sub1:sub1}

Initial conditions, leading to convergence are said to lie inside 
the solver's \emph{basin of attraction}. Inspecting \cref{sec2:fig4} 
this basin can be divided (roughly) into two main regions : the 
initial transition zone/region, $\mathcal{B_T}$, where the solver 
does not really converge but traverses the state-space until a suitable point 
is reached. Once this happens the solver 
enters the convergence zone/region, $\mathcal{B_C}$. 
Denote the iteration step where the solver state enters $\mathcal{B_C}$, or 
leaves $\mathcal{B_T}$, as $l_0 \in \mathbb{Z}_{\geq 0}$. The markers 
$\mathcal{B_T},\mathcal{B_C},l_0$ are illustrated 
in \cref{sec2:fig4}.(b), for $\Delta_f^{\text{\tiny init}}\sim 12\%$ (red curves), 
where the solver is seen to enter the convergence zone around $l=l_0=11$. Obviously, 
the goal is to enter $\mathcal{B_C}$ as quickly as possible, 
or equivalently, minimize the number of steps inside $\mathcal{B_T}$. 
Generally, $l_0$ should decrease w/ 
$\Delta_f^{\text{\tiny init}}$, as this will move the solver initial state 
towards a perfect PSS frequency guess (see \cref{sec2:eq3,sec2:eq1}). 
Thus, we expect to observe, $l_0 \to 0$, for, $\Delta_f^{\text{\tiny init}} \to 0$. 
Inspecting \cref{sec2:fig4}.(b), this trend can indeed be observed, \emph{i.e.} 
improving the frequency guess leads to quicker PSS convergence. Empirical rules,
of the type discussed above, however only describe a trend and do not guarantee 
that the relation always hold. \Cref{sec2:fig4}.(a) illustrates this issue as 
circuits \texttt{OSC.\#2-3} 
display as much smaller $l_0$ (reaches $\mathcal{B}_C$ faster) 
compared to \texttt{OSC.\#1} for identical initial conditions. 
It should be possible to advance the robustness of the \texttt{QUCS-COPEN} 
PSS algorithm by introducing 
\emph{e.g.} multiple-shooting solver methods\cite{Parkhurst2002} 
(see also \cref{sec1:sub1}) and implementing global \& local solver dampening techniques\cite{Kakizaki1985}. 
It will furthermore be interesting to look into employing 
Krylov (matrix-free) linear solvers \cite{Kundert1995,Kundert1997}, 
something not currently available in the \texttt{QUCS} package,
which should significantly increase efficiency, 
especially for large circuits. 
The ideas mentioned above will all be considered 
for future work on the \texttt{QUCS-COPEN} project.

\subsubsection{The Convergence Rate Inside $\mathcal{B}_C$.}
\label{sec2:sub1:sub2}

We seek to quantify the convergence order \& rate for the \texttt{QUCS-COPEN} 
PSS module experiments in \cref{sec2:fig4} inside convergence zone $\mathcal{B}_C$. 
The basic concepts required for this discussion are reviewed 
in \cref{app1:sec1}. For our purposes, 3 additional sequences, w/ 3 convergence orders 
are plotted in \cref{sec2:fig4} (dotted curves, see caption for details). 
Comparing the convergence curves w/ these added test-sequences, 
the following observations \emph{w.r.t} the rate of convergence inside 
$\mathcal{B}_C$ can be stated.

\begin{enumerate}
\item the solver never obtains the ideal Newton solver quadratic convergence ($\sigma =2$).
\item the convergence order inside, $\mathcal{B}_C$, is larger than 1, \emph{i.e.} 
faster than linear-convergence. 
\item from inspection of the various curves in \cref{sec2:fig4}, a rough estimate of the 
convergence order inside, $\mathcal{B}_C$, could be around $\sigma = 1.5$ (compare 
with brown dotted curve).
\item the linear-convergence test-sequence (purple dotted) represents a rough 
tangential to the convergence curves around the PSS 
convergence point, $\epsilon_{c} = \texttt{EpsMax}$. Hence, in a small region 
around $\epsilon_c$ the error sequence measures, $\epsilon(l),\Delta_f(l)$, 
defined in \cref{sec2:eq1,sec2:eq2}, can be approximated as linear sequences w/ 
convergence rate $\mu_L =2$. \label{sec2:list1:muL}
\end{enumerate}

Let, $\sigma_{\text{\tiny QC}}$, denote the convergence order of the 
\texttt{QUCS-COPEN} PSS module, it then follows from the remarks 
that, $\sigma_{\text{\tiny QC}} \sim 1.5$, inside $\mathcal{B}_C$. 
Furthermore, from \cref{sec2:list1:muL} in the above list, since the linear approximation 
close to convergence, $\epsilon \simeq \epsilon_c$, has convergence-rate 
$\mu_L = 2$, it follows that the precision of the solution, as measured by the closeness 
of, $\epsilon,\Delta_f$, to zero, increases 
two decimal digits for each iteration step 
in this area (see \cref{app1:sec1} for details).
\par
The \texttt{QUCS-COPEN} PSS module is not able 
to attain the ideal Newton-solver quadratic convergence. However, 
this should not be too surprising. This quadratic convergence rule, 
often quoted, is an idealized theoretical result and does 
not take into account the various issues involved with implementing a real-world 
numerical PSS solver, such as \emph{e.g.} operating w/ finite number precision, 
non-ideal linear matrix solvers, considering a nonlinear map which, implicitly, an involves a 
IVP which brings a whole new set of issues (\emph{e.g.} truncation error, internal linear solver errors) 
\emph{etc.} This is however not to say that there is no room for improvement, 
as this is almost certainly the case. We intend to explore various 
schemes for advancing the PSS tool \emph{w.r.t.} 
convergence rate, including, implementing the ideas discussed above in \cref{sec2:sub1:sub1}.

\section{Conclusion \& Future Work.}

\label{sec3}

The paper documents the work done implementing a 
periodic steady-state analysis module, based around a single-shooting/Newton-solver 
kernel algorithm, into the novel \texttt{QUCS-COPEN} simulator engine; a clone 
the \texttt{QUCS} and \texttt{QUCS-S} software projects. 
The work advances the state of the \texttt{QUCS} project as 
this represents, to our knowledge, the first time such 
a tool, applicable to autonomous circuits, has been developed for this environment. 
The PSS module 
was validated through a comparison-study w/ the commercial 
Keysight-ADS\textsuperscript{\textcopyright} EDA program and 
the convergence properties were analyzed w/ proposals and 
ideas for improving robustness and efficiency of the algorithm. 
Future work will focus on increasing the 
reliability \& efficiency of the tool by exploring more robust 
multiple-shooting solvers and using more elaborate solver dampening schemes 
to ensure convergence even for weak initial conditions. We furthermore intend 
to investigate the possible use of Krylov (matrix-free) linear-solvers which should 
increase the performance of the engine significantly, especially for larger circuits. 
The long-term goal of the work, described herein, 
is the development of a novel (coupled)-oscillator 
phase-noise analysis module to be integrated into the \texttt{QUCS-COPEN} distribution. 
This module builds on top of the steady-state tool described herein, and it 
is the topic of the second part of this paper-series.

\section*{Acknowledgment}

The authors gratefully acknowledge partial financial support by
German Research Foundation (DFG) (grant no. KR1016/17-1).

\appendix

\section{Brief Summary of Convergence Order \& Rate.}
\label{app1:sec1}

An iterated sequence, $f_k$, which converges asymptotically towards zero, 
$\lim_{k \to \infty} f_k = 0$, is said to have asymptotic 
convergence order \& rate, $\sigma,\mu$, if 
$\lim_{k\to \infty}|f_{k+1}|/|f_k|^\sigma = \mu$. Here $\sigma = 1$ 
is called linear-convergence whereas $\sigma=2$ refers to 
quadratic-convergence and so on. Note that, $\sigma = 1$, is referred to as 
"linear" convergence since a sequence, plotted w/ 
a log-linear axis configuration, will produce a straight line. Hence, 
"linear" convergence, in this context, actually implies 
exponential convergence curve whereas, $\sigma > 1$, denotes convergence 
faster than exponential. The Newton-Raphson (NR) solver, upon which 
the PSS algorithm discussed herein is based on, is known 
to converge w/ at-least quadratic order, in small 
open region containing the root. For linear convergence order, $\sigma = 1$, 
it is usual practice to introduce the convergence rate as 
$\mu_{L} = -\log_{10}(\mu)$ which expresses the number of 
decimal digits the solution improves, per iteration step. Thus, let $f_k$ be a 
sequence w/ linear convergence towards 0 and let $\mu_L = m$. 
In this case, the error, $e_k = |f_k|$, moves $m$ decimal digits closer 
to zero for each iteration step.

\end{document}